\documentstyle[12pt,psfig]{article}

\begin{document}

\title{How sensitive are various $NN$ observables to changes in
the $\pi NN$ coupling constant?~\footnote{Invited talk 
presented at the Workshop on 
{\it Critical Points in the Determination of the Pion-Nucleon 
Coupling Constant}, Uppsala (Sweden), June 7-8, 1999.}}

\author{R. Machleidt
\\Department of Physics, University of Idaho, 
\\Moscow, Idaho 83844, U.\ S.\ A.}

\maketitle

\begin{abstract}
The deuteron, $NN$ analyzing powers $A_y$, and the singlet scattering
length show great sensitivity to the $\pi NN$ coupling constant $g_\pi$.
While the $pp$ $A_y$ data favor 
$g^2_\pi/4\pi\leq 13.6$, 
the $np$ $A_y$ data and the deuteron quadrupole moment imply
$g^2_\pi/4\pi\geq 14.0$. 
The two diverging values could be reconciled by the assumption
of (substantial) charge-splitting of $g_\pi$.
However, the established theoretical explanation of the charge-dependence
of the $^1S_0$ scattering length (based upon pion mass splitting)
is very sensitive to a difference between $g_{\pi^0}$ and $g_{\pi^\pm}$
and rules out any substantial charge-splitting of $g_\pi$.
Thus, there are real and large discrepancies between
the values for $g_\pi$ extracted from different $NN$ observables.
Future work that could resolve the problems is suggested.
\end{abstract}

\newpage

\section{Introduction}
In this contribution, I will focus on the deuteron, $NN$ analyzing
powers, and the singlet scattering length. Other $NN$ observables
with sensitivity to the $\pi NN$ coupling constant are
the spin transfer coefficients $D_{NN'}$, $D_{LL'}$,
and $D_{SS'}$ which are discussed in the contribution
by Scott Wissink and in Ref,~\cite{BM95}.

\section{The deuteron}
In the 1980's, Torleif Ericson pointed out repeatedly~\cite{ER83}
that the deuteron is the most convincing manifestation of the
`existence' of the pion in nuclear physics.
Historically, it was the non-vanishing
quadrupole moment of the deuteron that provided the first evidence
for a nuclear tensor force, which is created by the pion.
Also, while the theoretical explanation of $NN$ scattering
observable requires, in general, to take several mesons into account,
the deuteron can be described by the pion alone (together
with a semi-soft $\pi NN$ form factor).
Thus, there are good physics reasons why the deuteron should
show a great deal of sensitivity to the $\pi NN$ coupling constant, $g_\pi$.

The crucial deuteron properties to consider are the quadrupole moment,
$Q$, and the asymptotic D/S state ratio, $\eta$.
The sensitivity of both quantities to $g_\pi$ is demonstrated in 
Table~I.  The calculations are based upon the most recent Bonn potential
(`CD-Bonn'~\cite{MSS96}) which belongs to the new generation
of high-precision $NN$ potentials that fit the $NN$ data below
350 MeV with a `perfect' $\chi^2$/datum of about one.
The numbers in Table~I are an update of earlier calculations
of this kind~\cite{MS91,ML93} in which older $NN$ potentials
were applied. There are no substantial differences in the results
as compared to the earlier investigations.

For meaningful predictions, it is important that all deuteron models
considered are realistic. This requires that besides the deuteron
binding energy (that is accurately reproduced by all models of
Table~I) also other empirically well-known quantities 
are correctly predicted, like
the deuteron radius, $r_d$,  and the triplet effective range
parameters, $a_t$ and $r_t$. As it turns out,
the latter quantities are closely related to the asymptotic S-state
of the deuteron, $A_S$, which itself is not an observable.
However, it has been shown~\cite{ER83} that for realistic values
of $r_d$, $a_t$, and $r_t$, the asymptotic S-state of the deuteron
comes out to be in the range $A_S=0.8845\pm 0.0008$ fm$^{-1/2}$.
Thus, $A_S$ plays the role of 
an important control number that tells us if a deuteron
model is realistic or not. As can be seen from Table~I, 
all our models pass the test.

Model A of Table~I uses the currently fashionable value
for the $\pi NN$ coupling constant
$g^2_\pi/4\pi = 13.6$ which clearly underpredicts $Q$
while $\eta$ is fine. 
One could now try to fix the problem with $Q$ 
by using a weaker
$\rho$-meson tensor-coupling to the nucleon, $f_\rho$.
It is customary to state the strength of this coupling in terms of the
tensor-to-vector ratio of the $\rho$ coupling constants,
$\kappa_\rho\equiv f_\rho/g_\rho$. Model A uses the `large' value
$\kappa_\rho=6.1$ recommended by Hoehler and Pietarinen~\cite{HP75}. 
Alternatively, one may try the value implied by the vector-meson
dominance model for the electromagnetic form factor of the 
nucleon~\cite{Sak69}
which is $\kappa_\rho=3.7$. This is done in our Model B which shows
the desired improvement of $Q$. However, a realistic model for the
$NN$ interaction must not only describe the deuteron but also
$NN$ scattering. As discussed in detail in Ref.~\cite{BM94},
the small $\kappa_\rho$ cannot reproduce the $\epsilon_1$ mixing
parameter correctly and, in addition, there are serious problems with
the $^3P_J$ phase shifts, particularly, the $^3P_0$ (cf.\ lower
part of Table~I and Fig.~1). Therefore, Model B is unrealistic and must
be discarded.

\begin{table}[t]
\footnotesize
{Table~I. Important coupling constants and the predictions for the deuteron
and some $pp$ phase shifts for five models discussed in the text.}
\begin{center}
\begin{tabular}{ccccccc}
\hline\hline
 \hspace*{2.4cm} 
                   &~A~&~B~&~C~&~D~&~E~
   &\hspace*{.4cm}Empirical\hspace*{.4cm}\\
\hline\hline
\multicolumn{7}{c}{\bf Important coupling constants}\\
~$g^2_{\pi^0}/4\pi$~&~13.6~&~13.6~&~14.0~&~14.4~&~13.6~&~\\
~$g^2_{\pi^\pm}/4\pi$~&~13.6~&~13.6~&~14.0~&~14.4~&~14.4~&~\\
~$\kappa_\rho$~&~6.1~&~3.7~&~6.1~&~6.1~&~6.1~&~\\
\hline
\multicolumn{7}{c} {\bf The deuteron}\\
~$Q$ 
(fm$^2$)&~0.270~&~0.278~&~0.276~&~0.282~&~0.278~&~0.276(2)$^a$~\\
~$\eta$&~0.0255~&~0.0261~&~0.0262~&~0.0268~&~0.0264
&~0.0256(4)$^b$~\\
$A_S$ (fm$^{-1/2}$)&0.8845&0.8842&0.8845&0.8845&0.8847&0.8845(8)$^c$\\
~$P_D$ (\%)&~4.83~&~5.60~&~5.11~&~5.38~&~5.20~&~--\\
\hline
\multicolumn{7}{c}{{\bf \mbox{\boldmath $^3P_0$ $pp$} phase shifts} (deg)}\\
10 MeV & 3.726 & 4.050 & 3.881 & 4.039 & 3.726  & 3.729(17)$^d$\\  
25 MeV & 8.588 & 9.774 & 8.981 & 9.384 & 8.588  & 8.575(53)$^d$\\
50 MeV &11.564&14.070&12.158& 12.763&11.564 & 11.47(9)$^d$ \\
\hline\hline
\end{tabular}
\end{center}
\footnotesize
$^a)$ Corrected for meson-exchange currents and relativity~\cite{foot1}.
\\
$^b)$ Ref.~\cite{RK90}.
\\
$^c)$ Ref.~\cite{ER83}.
\\
$^d)$ Nijmegen $pp$ multi-energy phase shift analysis~\cite{Sto93}.
\end{table}

The only parameters left to improve $Q$ are $g_\pi$ and the cutoff
mass, $\Lambda_\pi$, that is used to parametrize the $\pi NN$ form factor
(cf.\ Eq.~(6) below).
Similar, to the $\rho$ meson, $\Lambda_\pi$
is heavily constrained by $NN$ phase parameters, particularly,
$\epsilon_1$. The accurate reproduction of $\epsilon_1$
as determined in the Nijmegen $np$ multi-energy phase shift 
analysis~\cite{Sto93} essentially leaves no room for variations
of $\Lambda_\pi$ once the $\rho$ meson parameters are fixed.

Thus, we are finally left with only one parameter to fix the
$Q$ problem, namely $g_\pi$. As it turns out, for relatively
small changes of 
$g^2_\pi/4\pi$
there is a linear relationship, as demonstrated in Table~I
by the predictions of Model A, C and D which use
$g^2_\pi/4\pi=13.6$, 14.0, and 14.4, respectively.
Consistent with our earlier studies~\cite{MS91,ML93},
we find that 
$g^2_\pi/4\pi\geq14.0$
is needed to correctly reproduce $Q$.

\begin{figure}[t]
\psfig{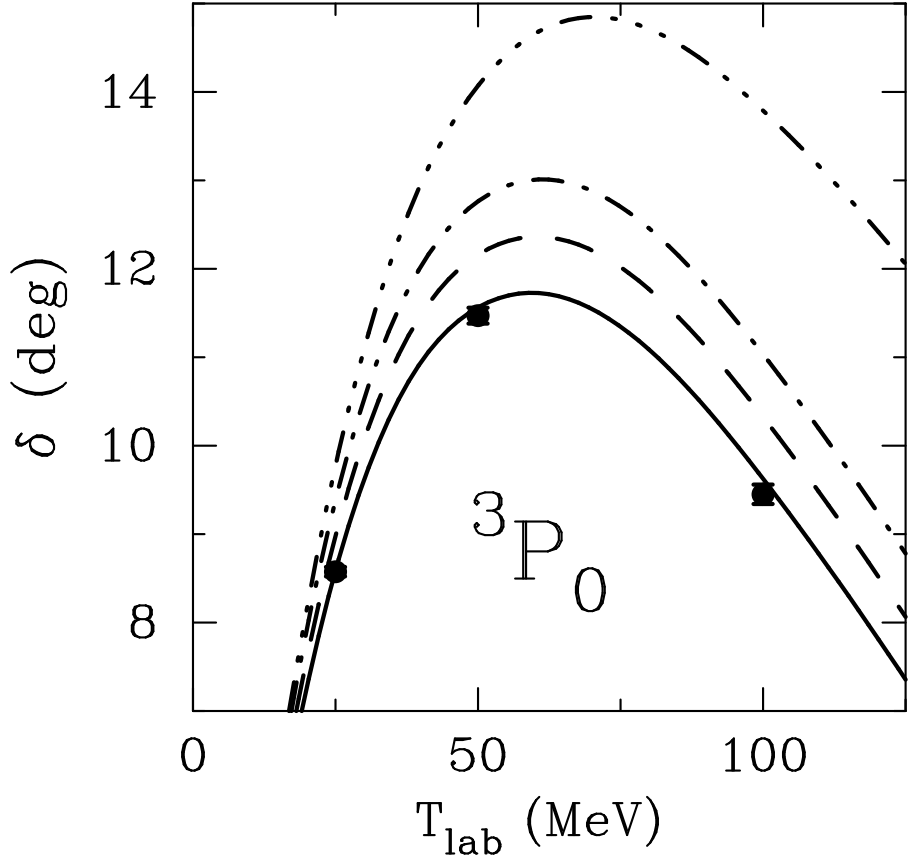}
\vspace*{1cm}\caption{$^3P_0$ phase shifts of proton-proton scattering
as predicted by Model A and E ($g^2_{\pi^0}/4\pi=13.6$, 
solid line), 
B ($\kappa_\rho=3.7$, 
dash-3dot), C ($g^2_{\pi^0}/4\pi=14.0$, 
dashed), and D ($g^2_{\pi^0}/4\pi=14.4$, 
dash-dot)~\cite{foot3}.
The solid dots represent the Nijmegen $pp$ multi-energy phase shift
analysis~\cite{Sto93}.}
\end{figure}

However, a pion coupling with
$g^2_\pi/4\pi\geq14.0$
creates problems for the $^3P_0$ phase shifts which are
predicted too large at low energy (cf.\ lower part of Table~I
and Fig.~1).
Now, a one-boson-exchange (OBE) model for the $NN$ interaction
includes several parameters (about one dozen in total).
One may therefore try to improve the
$^3P_0$ by readjusting some of the other model parameters.
The vector mesons ($\rho$ and $\omega$) have a strong impact on the
$^3P_0$ (and the other $P$ waves).
However, due to their heavy masses, they are more effective at high
energies than at low ones. Therefore, $\rho$ and $\omega$ 
may produce large changes of the $^3P_0$ phase shifts
in the range 200-300 MeV, with
little improvement at low energies. The bottom line is that
in spite of the large number of parameters in the model,
there is no way to fix the $^3P_0$ at low energies. In this
particular partial wave, 
the pion coupling constant is the only effective parameter, 
at energies below 100 MeV.
The $pp$ phase
shifts of the Nijmegen analysis~\cite{Sto93} as well as 
the $pp$ phases produced by the VPI group~\cite{SAID} require
$g^2_\pi/4\pi\leq13.6$.

Notice that this finding is in clear contradiction to our conclusion
from the deuteron $Q$.

There appears to be a way to resolve this problem.
One may assume that the neutral pion, $\pi^0$, couples to the
nucleon with a slightly different strength than the charged
pions, $\pi^\pm$. This assumption of a charge-splitting of
the $\pi NN$ coupling constant is made in our Model E where
we use
$g^2_{\pi^0}/4\pi=13.6$ and
$g^2_{\pi^\pm}/4\pi=14.4$.
This combination reproduces the $pp$ $^3P_0$ phase shifts at low
energy well~\cite{foot3}
and creates a sufficiently large deuteron $Q$.

\begin{table}[t]
{Table~II. $\chi^2$/datum for the fit of the world $pp$ $A_y$ data below 350 MeV
(subdivided into three energy ranges) using different values
of the $\pi NN$ coupling constant~\cite{foot3}.}
\begin{center}
\begin{tabular}{ccccc}
\hline\hline
\hspace*{7.0cm} 
  & \multicolumn{4}{c}{\bf Coupling constant \mbox{\boldmath $g^2_{\pi^0}/4\pi$}}\\
Energy range (\# of data) & 13.2 & 13.6 & 14.0 & 14.4 \\
   &   & A & C & D \\
\hline
0--17 MeV (45 data) & 0.84 & 1.43 & 2.71 & 4.66 \\
17--125 MeV (148 data) & 1.05 & 1.06 & 1.54 & 2.45 \\
125--350 MeV (624 data) & 1.24 & 1.22 & 1.26 & 1.34 \\
\hline\hline
\end{tabular}
\end{center}
\end{table}

\section{Analyzing powers}

In our above considerations, 
some $pp$ phase shifts played
an important role. 
In principle, phase shifts are
nothing else but an alternative representation of data.
Thus, one may as well use the data directly. 
Since the days of Gammel and Thaler~\cite{GT57}, it is well-known
that the triplet $P$-wave phase shifts (which we focused on, above)
are fixed essentially by the $NN$
analyzing powers, $A_y$. Therefore, we will now take a look
at $A_y$ data and compare them directly with model predictions.

In Fig.~2, we show high-precision $pp$ $A_y$ data at 9.85 MeV
from Wisconsin~\cite{Bar82}.
The theoretical curves shown are obtained with
$g^2_{\pi^0}/4\pi=13.2$ (dotted), 13.6 (solid), and 14.4 (dash-dot)
and fit the data with a $\chi^2$/datum of
0.98, 2.02, and 9.05, respectively.
Clearly, a small coupling constant around 13.2 is favored.
Since a single data set is not a firm basis, we have looked into
all $pp$ $A_y$ data in the energy range 0--350 MeV.
Our results are presented in Table~II where we give the $\chi^2$/datum
for the fit of the world $pp$ $A_y$ data below 350 MeV
(subdivided into three energy ranges) for various choices
of the neutral $\pi NN$ coupling constant.
It is seen that the $pp$ $A_y$ data at low energy, particularly
in the energy range 0--17 MeV, are very sensitive to the
$\pi NN$ coupling constant.
A value $g^2_{\pi^0}/4\pi \leq 13.6$ is clearly preferred, consistent
with what we extracted from the single data set at 9.85 MeV as well as
from $^3P_0$ phase shifts
in the previous section (cf.\ Fig.~1).

\begin{figure}[t]
\psfig{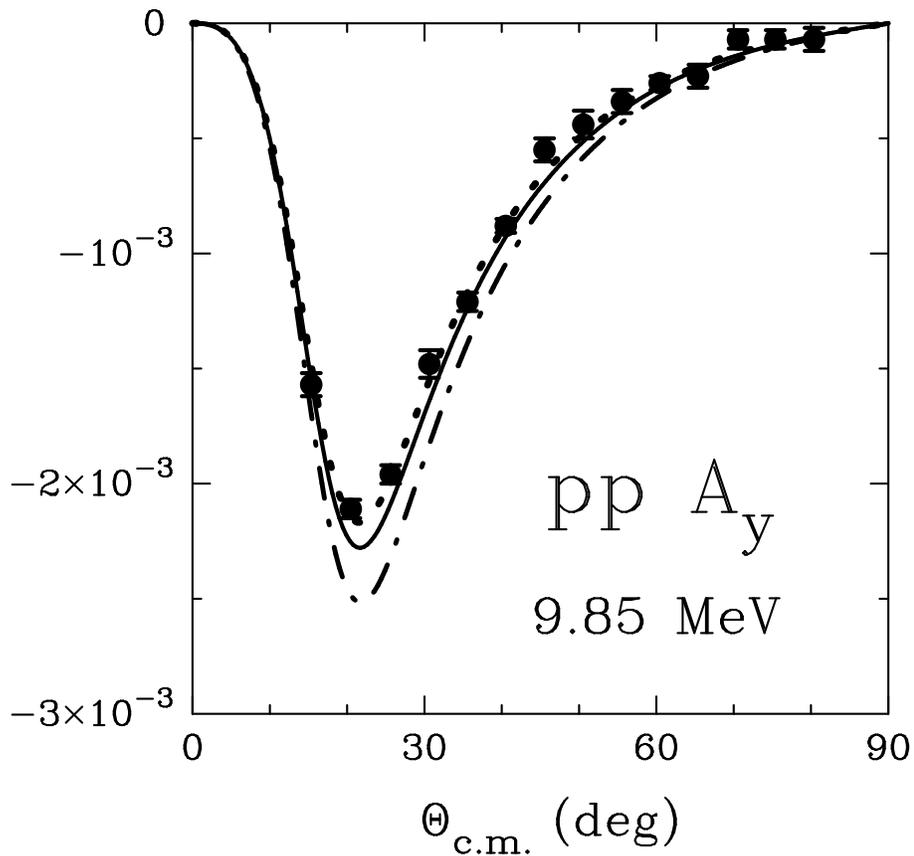}
\vspace*{1cm}\caption{The proton-proton analyzing power $A_y$
at 9.85 MeV.
The theoretical curves are calculated with
$g^2_{\pi^0}/4\pi=13.2$ (dotted), 
13.6 (solid, Model~A), and 14.4 (dash-dot, Model~D)
and fit the data with a $\chi^2$/datum of
0.98, 2.02, and 9.05, respectively.
The solid dots represent the data taken at
Wisconsin~\cite{Bar82}.}
\end{figure}

Next, we look into the $np$ $A_y$ data.
A single sample is shown in Fig.~3,
the $np$ $A_y$ data at 12 MeV from TUNL~\cite{Wei92}.
Predictions are shown for Model A (solid line), D (dash-dot),
and E (dash-triple-dot). The charge-splitting Model E
fits the data best with a $\chi^2$/datum of 1.00 (cf.\ Table~III).
We have also considered the entire $np$ $A_y$ data measured
by the TUNL group~\cite{Wei92} in the energy range 7.6--18.5 MeV
(31 data) as well as 
the world $np$ $A_y$ data
in the energy ranges 0--17 MeV (120 data).
It is seen that there is some sensitivity to the $\pi NN$
coupling constant in this energy range, while there is little
sensitivity at energies above 17 MeV (cf.\ Table~III).

Consistent with the trend seen in the 12 MeV data, 
the larger data sets below 17 MeV show a clear preference for
a coupling constant around 14.4 if there is no charge splitting
of $g_\pi$. 
This implies that without charge-splitting
it is impossible to obtain
an optimal fit of the $pp$ and $np$
$A_y$ data. To achieve this best fit charge-splitting is needed,
like $g^2_{\pi^0}/4\pi=13.6$
and $g^2_{\pi^\pm}/4\pi=14.0$
as considered in column 5 of Table~III.
The drastic charge-splitting of Model E is not favored by the
more comprehensive $np$ $A_y$ data sets.

\begin{table}[t]
\footnotesize
{Table~III. $\chi^2$/datum for the fit of various sets of $np$ $A_y$ data
using different values
for the $\pi NN$ coupling constants.}
\begin{center}
\begin{tabular}{cccccc}
\hline\hline
  & \multicolumn{5}{c}{\bf Coupling constants \mbox{\boldmath 
$g^2_{\pi^0}/4\pi; \; g^2_{\pi^\pm}/4\pi$}}\\
Energy, data set (\# of data) &~13.6; 13.6~&~14.0; 14.0~&~14.4; 14.4
 &~13.6; 14.0~&~13.6; 14.4~\\
   &~A~&~C~&~D~&   &~E~\\
\hline\hline
12 MeV \cite{Wei92} (9 data)
 & 2.81 & 2.27 & 1.79 & 1.53 & 1.00 \\
7.6--18.5 MeV \cite{Wei92} (31 data)
 & 1.89 & 1.56 & 1.29 & 1.28 & 1.32 \\
0--17 MeV world data (120)
 & 1.17 & 1.03 & 0.94 & 0.99 & 1.19 \\
\hline
17--50 MeV \cite{Wil84} (85 data)
 & 1.16 & 1.12 & 1.14 & 1.18 & 1.18 \\
17--125 MeV world data (416)
 & 0.89 & 0.89 & 0.91 & 0.91 & 0.94 \\
\hline\hline
\end{tabular}
\end{center}
\end{table}

The balance of the analysis of the $pp$ and $np$ $A_y$ data then is: 
$g^2_{\pi^0}/4\pi \leq 13.6$
and $g^2_{\pi^\pm}/4\pi \geq 14.0$.
Notice that this
splitting is consistent with our conclusions in Sect.~2.
Thus, we have now some indications for charge-splitting of $g_\pi$
from two very different observables, namely the deuteron quadrupole
moment and $np$ analyzing powers.

Therefore, it is worthwhile to look deeper into the issue
of charge-splitting of the $\pi NN$ coupling constant.
Unfortunately, there are severe problems with any substantial
charge-splitting---for two reasons.
First, theoretical work~\cite{Hwa98} on isospin symmetry breaking
of the $\pi NN$ coupling constant based upon QCD sum rules comes
up with a splitting of less than 0.5\% for $g_\pi^2$ and, thus,
cannot explain the large charge splitting indicated above.
Second, a problem occurs with the conventional
explanation of the charge-dependence of the singlet
scattering length, which we will discuss in the next section.

\begin{figure}[t]
\psfig{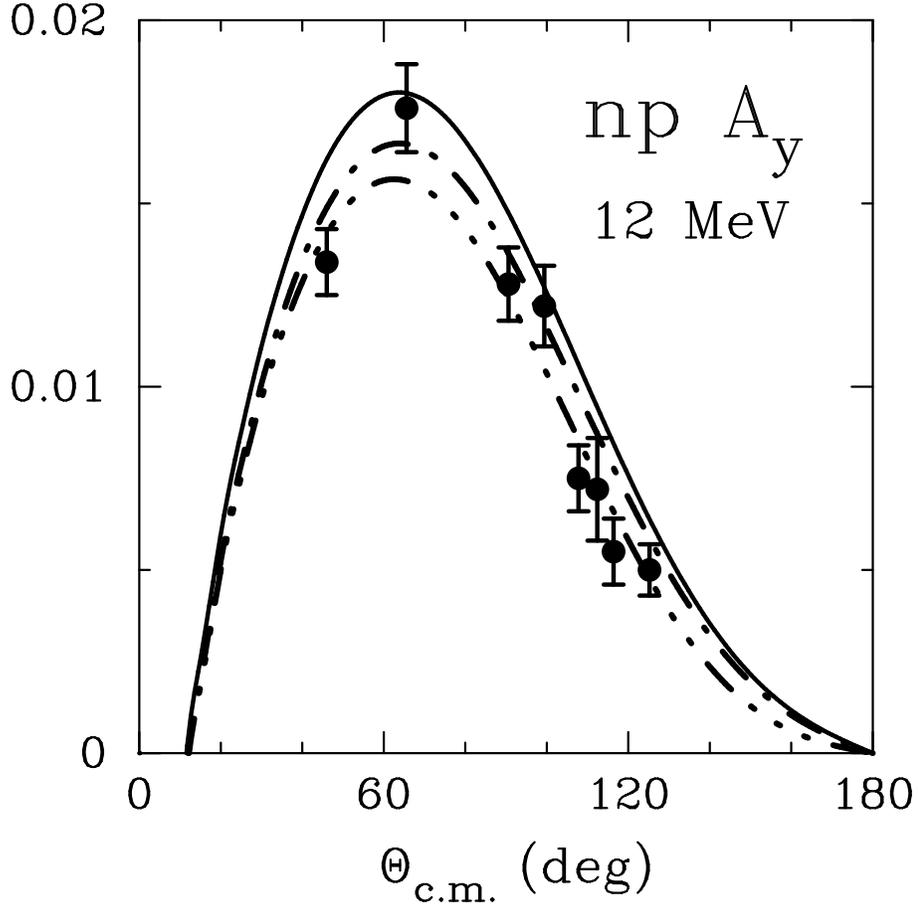}
\vspace*{1cm}\caption{The neutron-proton analyzing power $A_y$
at 12 MeV.
The theoretical curves are calculated with
$g^2_{\pi^0}/4\pi
=g^2_{\pi^\pm}/4\pi
=13.6$ (solid line, Model A), 
$g^2_{\pi^0}/4\pi
=g^2_{\pi^\pm}/4\pi
=14.4$ (dash-dot, Model D), and the charge-splitting
$g^2_{\pi^0}/4\pi=13.6,\;
g^2_{\pi^\pm}/4\pi=14.4$ (dash-3dot, Model E). 
The solid dots represent the data taken at
TUNL~\cite{Wei92}.}
\end{figure}

\section{Charge-dependence of the singlet scattering length and
charge-dependence of the pion coupling constant}
The ultimate purpose of this Section is to show in detail how 
charge-splitting of the $\pi NN$ coupling constant
affects the charge-dependence of the $^1S_0$ scattering
length. It will turn out that the charge-splitting of $g_\pi$
suggested in Sect.~2  and 3 causes a disaster for our established
understanding of the charge-dependence of the singlet
sattering length. 

To set the stage properly, I will first summarize the
established empirical and theoretical facts about
the charge-dependence of the nuclear force.

The equality between 
proton-proton ($pp$) [or neutron-neutron ($nn$)] and neutron-proton ($np$)
nuclear interactions is known as charge independence---a symmetry that
is slightly broken.  This is seen most clearly
in the $^1S_0$ nucleon-nucleon scattering lengths. 
The latest empirical values for the singlet scattering length $a$ 
and effective range $r$ are~\cite{MNS90,MO95}:
\begin{equation}
\begin{array}{lll}
a^N_{pp}=-17.3\pm 0.4 \mbox{ fm}, &\hspace*{2.5cm}
                                     & r^N_{pp}=2.85\pm 0.04 \mbox{ fm},\\
a^N_{nn}=-18.8\pm 0.3 \mbox{ fm}, && r^N_{nn} = 2.75\pm 0.11 \mbox{ fm},\\
a_{np}=-23.75\pm 0.01 \mbox{ fm}, && r_{np}=2.75\pm 0.05 \mbox{ fm}.
\end{array}
\end{equation}
The values given for $pp$ and $nn$ 
scattering refer to the nuclear part of the interaction
as indicated by the superscript $N$.
Electromagnetic effects have been removed from the experimental
values, which is model dependent. The uncertainties
quoted for $a^N_{pp}$ and $r^N_{pp}$ are due to this model dependence.

It is useful to define the following averages:
\begin{eqnarray}
\bar{a}\equiv \frac12 (a^N_{pp} + a^N_{nn}) & =&  -18.05\pm 0.5 \mbox{ fm},\\
\bar{r}\equiv \frac12 (r^N_{pp} + r^N_{nn}) & =&  2.80\pm 0.12 \mbox{ fm}.
\end{eqnarray}
By definition, charge-independence breaking (CIB) is the difference between 
the average of $pp$ and $nn$, on the one hand, and $np$ on the other: 
\begin{eqnarray}
\Delta a_{CIB} \equiv
 \bar{a}
 - 
 a_{np}
 &=& 5.7\pm 0.5 \mbox{ fm},\\
\Delta r_{CIB} \equiv
 \bar{r}
 - 
 r_{np}
 &=& 0.05\pm 0.13 \mbox{ fm}.
\end{eqnarray}
Thus, the $NN$ singlet scattering length shows a clear signature
of CIB in strong interactions.

The current understanding is 
that the charge dependence of nuclear forces is due to
differences in the up and down quark masses and electromagnetic 
interactions. 
On a more phenomenological level, major causes of CIB are the
mass splittings of isovector mesons (particularly, $\pi$ and $\rho$)
and irreducible pion-photon exchanges.

It has been known for a long time that the difference between the charged and 
neutral pion masses in the one-pion-exchange (OPE) potential  accounts
for about 50\% of $\Delta a_{CIB}$. 
Based upon the Bonn meson-exchange model for the $NN$ 
interaction~\cite{MHE87}, also multiple pion exchanges have been taken
into account. Including these interactions, about
80\% of the empirical $\Delta a_{CIB}$ can be explained~\cite{CM86,LM98}.
Ericson and Miller~\cite{EM83} obtained a very similar result using the
meson-exchange model of Partovi and Lomon~\cite{PL70}.

The CIB effect from OPE can be understood as follows.
In nonrelativistic approximation~\cite{foot2} and disregarding isospin
factors, OPE is given by
\begin{equation}
V_{1\pi}(g_\pi, m_\pi)  =  -\frac{g_{\pi}^{2}}{4M^{2}}
 \frac{({\mbox {\boldmath $\sigma$}}_{1} \cdot {\bf k})
       ({\mbox {\boldmath $\sigma$}}_{2} \cdot {\bf k})}
{m_{\pi}^2+{\bf k}^{2}}
\left( \frac{\Lambda^{2}_\pi-m_{\pi}^{2}}{\Lambda^{2}_\pi+{\bf k}^{2}}
\right)^n
\end{equation}
with $M$ the average nucleon mass, $m_\pi$ the pion mass, 
and {\bf k} the momentum transfer.
The above expression includes a form factor with 
cutoff mass $\Lambda_\pi$ and exponent $n$.

For $S=0$ and $T=1$, where $S$ and $T$ denote the total spin and isospin
of the two-nucleon system,
respectively, we have
\begin{equation}
V_{1\pi}^{01}(g_\pi, m_\pi)  =  
\frac{g_{\pi}^{2}}
{m_{\pi}^2+{\bf k}^{2}}
\frac{{\bf k}^2}
{4M^{2}}
\left( \frac{\Lambda^{2}-m_{\pi}^{2}}{\Lambda^{2}+{\bf k}^{2}}
\right)^n \; ,
\end{equation}
where the superscripts 01 refer to $ST$.
In the $^1S_0$ state, this potential expression is repulsive.
The charge-dependent OPE is then,
\begin{equation}
V_{1\pi}^{01}(pp)  =  
V_{1\pi}^{01}(g_{\pi^0}, m_{\pi^0})  
\end{equation}
for $pp$ scattering, and
\begin{equation}
V_{1\pi}^{01}(np)  =  
2 V_{1\pi}^{01}(g_{\pi^\pm}, m_{\pi^\pm})  
- V_{1\pi}^{01}(g_{\pi^0}, m_{\pi^0})  
\end{equation}
for $np$ scattering.

If we assume charge-independence of $g_\pi$ (i.~e., 
$g_{\pi^0}=g_{\pi^\pm}$), then all CIB comes from the charge
splitting of the pion mass, which is~\cite{PDG96}
\begin{eqnarray}
m_{\pi^0} & = & 134.976 \mbox{MeV,}\\
m_{\pi^\pm} & = & 139.570 \mbox{MeV.}
\end{eqnarray}

Since the pion mass appears in the denominator of OPE,
the smaller $\pi^0$-mass exchanged in $pp$ scattering
generates a larger (repulsive) potential in the $^1S_0$
state as compared to $np$ where also the larger $\pi^\pm$-mass
is involved. Moreover, the $\pi^0$-exchange in $np$
scattering carries a negative sign, 
which further weakens the $np$ OPE potential.
The bottom line is that the $pp$ potential is more repulsive
than the $np$ potential. The quantitative effect on
$\Delta a_{CIB}$
is about 3 fm (cf.\ Table IV).

\begin{table}[t]
\footnotesize
{Table~IV. Predictions for $\Delta a_{CIB}$ as defined in Eq.~(4)
in units of fm without and with the assumption of charge-dependence
of $g_\pi$.}
\begin{center}
\begin{tabular}{cccc}
\hline\hline
 & \multicolumn{2}{c}{\bf No charge-dependence of $g_\pi$} 
                                     & {\bf Charge-dependent $g_\pi$:} \\ 
 &                    &                           &$g^2_{\pi^0}/4\pi = 13.6$\\
 &Ericson \& Miller~\cite{EM83}&Li \& Machleidt~\cite{LM98}
                                            &$g^2_{\pi^\pm}/4\pi = 14.4$\\
\hline
$1\pi$ & 3.50 & 3.24 & --1.58 \\
$2\pi$ & 0.88 & 0.36 & --1.94 \\
$\pi\rho,\pi\sigma,\pi\omega$ & --- & 1.04 & --0.97 \\
\hline
Sum & 4.38 & 4.64 & --4.49 \\
\hline
Empirical & \multicolumn{3}{c}{$5.7\pm 0.5$}\\
\hline\hline
\end{tabular}
\end{center}
\end{table}

We now turn to the CIB created by the $2\pi$ exchange (TPE) contribution
to the $NN$ interaction. There are many diagrams that
contribute (see Ref.~\cite{LM98} for a complete overview).
For our qualitative discussion here, we pick the largest
of all $2\pi$ diagrams, namely, the box diagrams with
$N\Delta$ intermediate states, Fig.~4.
Disregarding isospin factors and using some drastic
approximations~\cite{foot2}, the amplitude for such a diagram is
\begin{equation}
V_{2\pi}(g_\pi, m_\pi)  =  -\frac{g_{\pi}^{4}}{16M^{4}}
\frac{72}{25} \int \frac{d^3p}{(2\pi)^3}
 \frac{[{\mbox {\boldmath $\sigma$}} \cdot {\bf k}
        {\mbox {\boldmath $S$}} \cdot {\bf k}]^2}
{(m_{\pi}^2+{\bf k}^{2})^2(E_p+E^\Delta_p-2E_q)}
\left( \frac{\Lambda^{2}-m_{\pi}^{2}}{\Lambda^{2}+{\bf k}^{2}}
\right)^{2n} \; ,
\end{equation}
where ${\bf k} = {\bf p} - {\bf q}$ with {\bf q}
the relative momentum in the initial and final state
(for simplicity, we are considering a diagonal matrix element); 
$E_p=\sqrt{M^2+{\bf p}^2}$ and $E^\Delta_p=\sqrt{M_\Delta^2+{\bf p}^2}$
with $M_\Delta=1232$ MeV the $\Delta$-isobar mass;
{\bf S} is the spin transition operator between nucleon and
$\Delta$. For the $\pi N\Delta$ coupling constant, $f_{\pi N\Delta}$,
the quark-model relationship
$f^2_{\pi N\Delta} = \frac{72}{25} f^2_{\pi NN}$ 
is used~\cite{MHE87}.

\begin{figure}[t]
\psfig{file=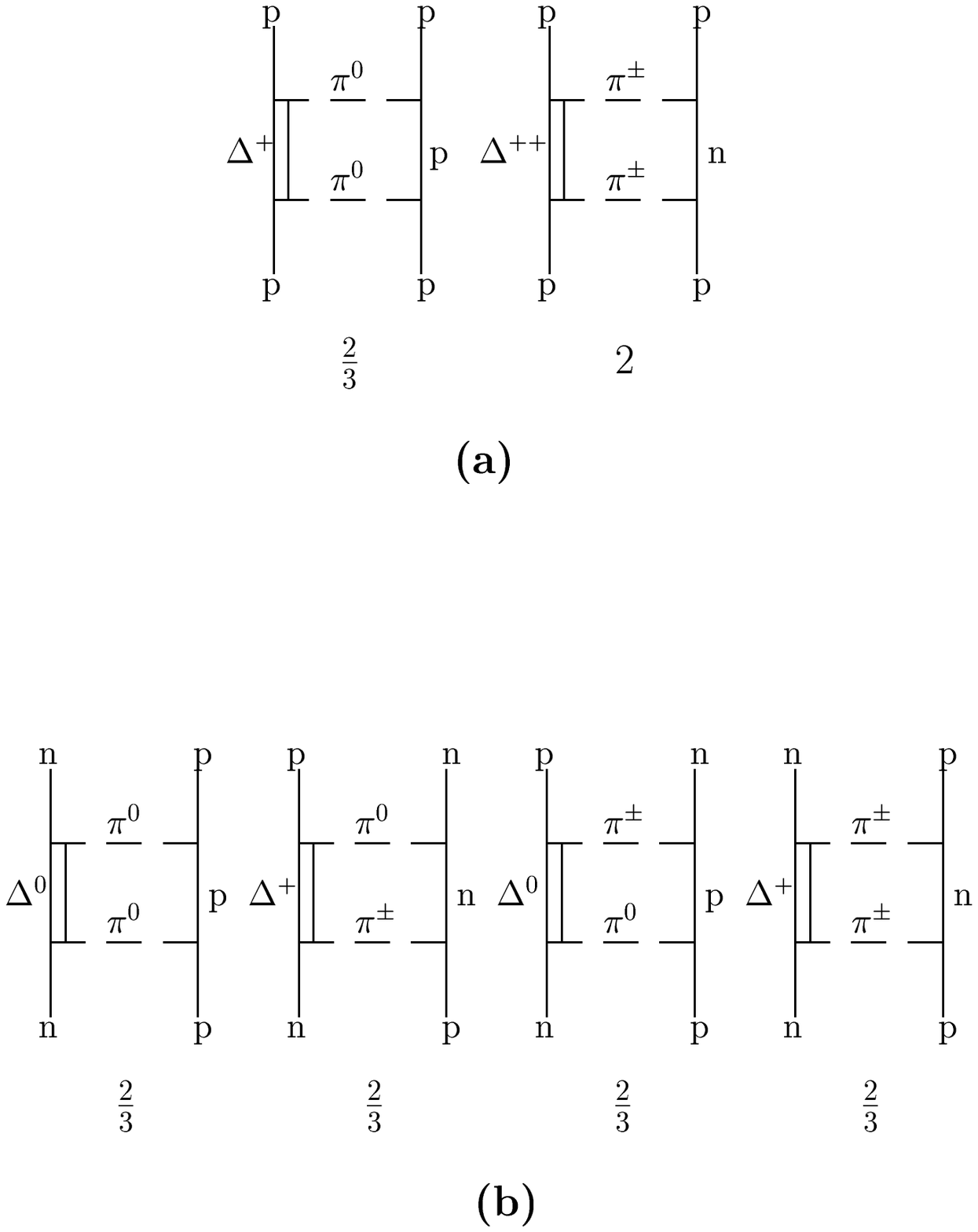,width=12cm}
\vspace*{-2cm}\caption{$2\pi$-exchange box diagrams
with $N\Delta$ intermediate states that contribute to 
(a) $pp$ and (b) $np$ scattering. 
The numbers below the diagrams are
the isospin factors.}
\end{figure}

For small momentum transfers {\bf k},
this attractive contribution is roughly proportional to
$m_\pi^{-4}$. Thus for TPE, the heavier pions will provide less attraction
than the lighter ones.
Charged and neutral pion exchanges
occur for $pp$ as well as for $np$, and it is important
to take the isospin factors carried by the various diagrams
into account. They are given in Fig.~4 below each diagram.
For $pp$ scattering, the diagram with 
double $\pi^\pm$ exchange
carries the largest factor, while 
double $\pi^\pm$ exchange
carries only a small relative weight in $np$ scattering.
Consequently, $pp$ scattering is less attractive than $np$
scattering which leads to an increase of
$\Delta a_{CIB}$ by 0.79 fm due to the diagrams of Fig.~4. 
The crossed diagrams of this type
reduce this result and including all $2\pi$ exchange diagrams 
one finds a total effect of 0.36 fm~\cite{LM98}.
Diagrams that go beyond $2\pi$ have also been investigated
and contribute another 1 fm.
In this way, pion-mass splitting explains about 80\% of 
$\Delta a_{CIB}$ (see Table IV for a summary).

Recall that our considerations in Sect.~2 sugested charge-splitting
of $g_\pi$, like
\begin{eqnarray}
g^2_{\pi^0}/4\pi & = & 13.6 \; , \\
g^2_{\pi^\pm}/4\pi & = & 14.4 \: ,
\end{eqnarray}
cf.\ Model E of Table~I.
We will now discuss how this charge-splitting of $g_\pi$ affects
$\Delta a_{CIB}$ (more details can be found in the original paper
Ref.~\cite{MB99}).

Accidentally, this splitting is---in relative terms---about the same as the
pion-mass splitting; that is
\begin{equation}
\frac{g_{\pi^0}}{m_{\pi^0}} \approx
\frac{g_{\pi^\pm}}{m_{\pi^\pm}}\; . 
\end{equation}
As discussed, for zero momentum transfer, we habe roughly
\begin{equation}
\mbox{OPE} \sim \left(\frac{g_\pi}{m_\pi}\right)^2
\end{equation}
and
\begin{equation}
\mbox{TPE} \sim \left(\frac{g_\pi}{m_\pi}\right)^4 \; ,
\end{equation}
which is not unexpected, anyhow.
On the level of this qualitative discussion, we can then predict that
any pionic charge-splitting 
satisfying Eq.~(15) will create no CIB from pion exchanges.
Consequently, a charge-splitting of $g_\pi$ as given in Eqs.~(13)
and (14) will wipe out our established explanation of CIB
of the $NN$ interaction.

We have also conducted accurate numerical calculations based upon the Bonn
meson-exchange model for the $NN$ interaction~\cite{MHE87}.
The details of these calculations are spelled out
in Ref.~\cite{LM98} where, however, no charge-splitting of $g_\pi$
was considered. Assuming the $g_\pi$ of 
Eqs.~(13) and (14), we obtain the $\Delta a_{CIB}$ 
predictions given in the last column of Table~IV.
It is seen that the results of an accurate calculation go even beyond
what the qualitative estimate suggested:
the conventional CIB prediction is not only reduced, it is reversed.
This is easily understood if one recalls that the pion mass appears
in the propagator $(m_\pi^2+{\bf k}^2)^{-1}$. Assuming an
average ${\bf k}^2\approx m^2_\pi$, the 7\% charge splitting of
$m^2_\pi$ will lead to only about a 3\% charge-dependent effect from
the propagator. Thus, if a 6\% charge-splitting of $g_\pi^2$ is used, 
this will not only override the pion-mass effect, it will reverse it.

Based upon this argument and on our numerical results, 
one can then estimate that
a charge-splitting of $g_\pi^2$ of only about 3\%
(e.~g., 
$g^2_{\pi^0}/4\pi = 13.6$ and $g^2_{\pi^\pm}/4\pi = 14.0$)
would erase all CIB prediction of the singlet scattering length
derived from pion mass splitting.

Besides pion mass splitting, we do not know of any other essential mechanism
to explain the charge-dependence of the singlet scattering length. 
Therefore, it is unlikely
that this mechnism is annihilated by a charge-splitting of $g_\pi$.
This may be taken as an indication that there is no significant
charge splitting of the $\pi NN$ coupling constant.

\section{Conclusions}

Several $NN$ observables can be identified that are very 
sensitive to the $\pi NN$ coupling constant, $g_\pi$.
They all carry the potential to determine
$g_\pi$ with high precision.

In particular, we have shown that the $pp$ $A_y$ data below
17 MeV are very sensitive to $g_\pi$ and imply a value
$g^2_\pi/4\pi \approx 13.2$.
The $np$ $A_y$ data below 17 MeV show moderate sensitivity
and the deuteron quadrupole moment shows great sensitivity to $g_\pi$;
both $np$ observables imply $g^2_\pi/4\pi\geq 14.0$.

The two different values may suggest a relatively large
charge-splitting of $g_\pi$. However, in Sect.~4, we have show that
a charge-splitting of this kind would completely erase our
established explanation of the charge-dependence of the
singlet scattering length. Since this is unlikely to be true,
we must discard the possibility of any substantial charge-splitting
of $g_\pi$.

The conclusion then is that we are faced with real and substantial
discrepancies between the values for $g_\pi$ based upon
different $NN$ observables.
The reason for this can only be that there are large,
unknown systematic errors in the data and/or large
uncertainties in the theoretical methods.
Our homework for the future is to find these errors and
eliminate them.

Another way to summarize the current cumbersome situation is to state
that, presently, any value between 13.2 and 14.4 is possible for
$g^2_\pi/4\pi$ depending on which $NN$ observable you pick.
If we want to pin down the value more tightly, then we are faced with
three possible scenarios:
\begin{itemize}
\item
$g_\pi$ is  small,
$g^2_\pi/4\pi\leq 13.6$:\\
The deuteron $\eta$ and $pp$ scattering
at low energies are described well; 
there are moderate problems with the $np$ $A_y$ data
below 17 MeV.
{\it The most serious problem is the deuteron $Q$.}
Meson-exchange current contributions (MEC) and relativistic
corrections for $Q$ of 0.016~fm$^2$ or more would solve the problem.
Present calculations predict about 0.010~fm$^2$ or less.
A serious reinvestigation of this issue is called for.
We note that an alternative solution of the problem with $Q$
is to introduce a heavy pion, $\pi'(1200)$. This possibility is discussed
in Ref.~\cite{ML93}.
\item
$g_\pi$ is large,
$g^2_\pi/4\pi\geq 14.0$:\\
The deuteron $Q$ is well reproduced, but $\eta$ is predicted too large
as compared to the most recent measurement by Rodning and Knutsen~\cite{RK90},
$\eta = 0.0256(4)$.
Note, however, that all earlier measurements of $\eta$ came up
with a larger value; for example, Borbely {\it et al.}~\cite{Bor89}
obtained $\eta = 0.0273(5)$. 
There are no objectively verifiable reasons
why the latter value should be less reliable than the former one.
The deuteron $\eta$ carries the potential of being the best observable to
determine $g_\pi$ (as pointed out repeatedly by Ericson~\cite{ER83}
in the 1980's); but the unsettled experimental situation
spoils it all. 
The $np$ $A_y$ data at low energy are described well.
{\it The most serious problem are the $pp$ $A_y$ data
below 100 MeV.}
\item
$g_\pi$ is `in the middle',
$13.6 \leq g^2_\pi/4\pi \leq 14.0$:\\
we have all of the above problems,
but in moderate form.
\end{itemize}

\noindent
{\bf Acknowledgement}

\noindent
This work was supported in part by the U.S.\ National
Science Foundation under Grant No.\ PHY-9603097.


\begin{thebibliography}{99}
\bibitem{foot1} From the experimental value for $Q$, 
which is 0.2859(3) fm$^2$~\cite{RV75,ER83}, we have subtracted 0.010(2) fm$^2$
to eliminate contributions obtained from meson-exchange currents
and relativistic corrections as calculated by Henning using the
Bonn potential~\cite{Hen93}. Note that our predictions
for $Q$ do not include the latter contributions.
\bibitem{foot2} For pedagogical reasons, we use simple, approximate
expressions when discussing the effects from pion exchange in a
qualitative manner. Note, however, that in our numerical
calculations we use relativistic time-ordered perturbation
theory in its full complexity and without approximations,
as outlined in all detail in the appendices of Ref.~\cite{MHE87}.
\end{thebibliography}
\end{document}